\documentclass{aa}

\usepackage{graphicx}
\usepackage{txfonts}
\usepackage[utf8]{inputenc}
\usepackage{lscape}
\usepackage{subcaption}
\usepackage{placeins}
\usepackage{siunitx}
\usepackage{natbib}
\usepackage{xcolor}
\usepackage{hyperref}
\usepackage{academicons}
\usepackage{orcidlink}
\usepackage{xcolor}
\bibpunct{(}{)}{;}{a}{}{,}
\newcommand{\comment}[1]{}

\begin{document}

\title{Resolved Kennicutt-Schmidt law in two strongly lensed star-forming galaxies at redshift 1}
\titlerunning{Resolved KS law in two strongly lensed SFGs at $z=1$}
\authorrunning{D. Nagy et al.}

\author{David Nagy\inst{1}\fnmsep\thanks{email: david.nagy@unige.ch} \and Miroslava Dessauges-Zavadsky\inst{1} \and Matteo Messa\inst{1,2} \and Johan Richard\inst{3} \and Jiayi Sun\inst{4,5}\orcidlink{0000-0003-0378-4667} \and Françoise Combes\inst{6} \and Yannick Eyholzer\inst{1}
          }

\institute{
        Département d’Astronomie, Université de Genève, Chemin Pegasi 51, 1290 Versoix, Switzerland
    \and
        Department of Astronomy, Stockholm University, AlbaNova University Centre, SE-106 91 Stockholm, Sweden
    \and
         Université Lyon, Université Lyon1, ENS de Lyon, CNRS, Centre de Recherche Astrophysique de Lyon UMR5574, Saint-Genis-Laval, France
    \and
        Department of Physics and Astronomy, McMaster University, 1280 Main Street West, Hamilton, ON L8S 4M1, Canada
    \and
        Canadian Institute for Theoretical Astrophysics (CITA), University of Toronto, 60 St George Street, Toronto, ON M5S 3H8, Canada
    \and
        LERMA, Observatoire de Paris, PSL Research Université, CNRS, Sorbonne Université, UPMC, Paris, France
        }


 
\abstract
    {We study the star formation rate (SFR) versus molecular gas mass ($M_\mathrm{mol}$) scaling relation from hundreds to thousands parsec scales in two strongly lensed galaxies at redshift $z\sim 1$, the Cosmic Snake and A521. We trace SFR using extinction-corrected rest-frame UV observations with the Hubble Space Telescope (HST), and $M_\mathrm{mol}$ using detections of the CO(4--3) line with the Atacama Large Millimetre/submillimetre Array (ALMA). The similar angular resolutions of our HST and ALMA observations of $0.15-\SI{0.2}{\arcsec}$ combined with magnifications reaching $\mu>20$ enable us to resolve structures in the galaxies of sizes lower than $\SI{100}{pc}$. These resolutions are close to those of nearby galaxies studies. This allows us to investigate for the first time the Kennicutt-Schmidt (KS) law (SFR-$M_\mathrm{mol}$ surface densities) at different spatial scales, from galactic scales to $\SI{\sim 100}{pc}$ scales, in galaxies at $z\sim 1$. At integrated scales we find that both galaxies satisfy the KS law defined by galaxies at redshifts between 1 and 2.5. We test the resolved KS (rKS) law in cells of sizes down to $\SI{200}{pc}$ in the two galaxies. We observe that this relationship generally holds in these $z\sim 1$ galaxies although its scatter increases significantly with decreasing spatial scales. We check the scale dependence of the spatial correlation between the surface densities of SFR and $M_\mathrm{mol}$ by focussing on apertures centred on individual star-forming regions and molecular clouds. We conclude that star-forming regions and molecular clouds become spatially de-correlated at $\lesssim \SI{1}{kpc}$ in the Cosmic Snake, whereas they appear de-correlated at all spatial scales (from $\SI{400}{pc}$ to $\SI{6}{kpc}$) in A521.}

\keywords{galaxies: high-redshift - galaxies: structure - gravitational lensing: strong - stars: formation}

\maketitle
%

\section{Introduction}

The star formation rate (SFR) and the total atomic (H\,{\scriptsize I}) and molecular (H$_2$) gas mass ($M_\mathrm{gas}$) of galaxies are closely related. Hydrogen being the primary fuel for star formation, its mass content is expected to correlate with SFR. A study of the SFR-$M_\mathrm{gas}$ relation by \citet{schmidt_rate_1959} revealed a clear correlation between the volume densities of SFR and $M_{\mathrm{gas}}$, and in \citet{schmidt_rate_1963} it was recast as a power law relationship between surface densities ($\Sigma$): $\Sigma \mathrm{SFR} = A \left (\Sigma M_\mathrm{gas}\right)^n$. \citet{kennicutt_star_1998} measured a power law index $n$ of the relation of $1.4\pm 0.15$ in local galaxies. H$_2$ is the gas phase in which the majority of star formation occurs, as it is the densest and coldest phase of the interstellar medium. A galaxy with a high H$_2$ mass ($M_\mathrm{mol}$) content is thus expected to form stars more efficiently. Therefore, the SFR-$M_\mathrm{mol}$ relation, commonly called the molecular Kennicutt-Schmidt (KS) law, has been extensively studied. It has also the form of a power law: $\Sigma \mathrm{SFR} = A \left (\Sigma M_\mathrm{mol}\right)^n$. Recent studies of the KS law report an index $n$ of $1.03\pm 0.08$ (e.g. \citealt{de_los_reyes_revisiting_2019}).

The surface densities in the KS law are integrated quantities measured on the whole galaxy. With the increasing availability of high resolution multiwavelength data for nearby galaxies, recent studies have been focusing on the investigation of the KS law at sub-galactic scales (\citealt{bigiel_star_2008,feldmann_how_2011,pessa_star_2021,leroy_molecular_2013,sun_star_2023}). A conclusion of these studies is that the resolved KS (rKS) law holds down to sub-kiloparsec spatial scales with a power law index around $1-1.1$, depending on the resolution. However, the scatter of the relation is expected to increase as the spatial scale decreases due to the statistical undersampling of the stellar IMF as well as time evolution of individual star-forming regions (e.g. \citealt{schruba_scale_2010,kruijssen_uncertainty_2018,pessa_star_2021}).

The molecular gas-to-SFR ratio, also called the molecular depletion time ($\tau_{\mathrm{dep}} = \Sigma M_\mathrm{mol}/\Sigma \mathrm{SFR}$), is the quantity that traces the time it would take for the molecular gas reservoir  to get consumed assuming a constant SFR. If stars are formed in giant molecular clouds (GMCs) for many dynamical times, or in other words if the star-forming process is in quasi-equilibrium at the scale of a single GMC, then the molecular gas and young stars are expected to correlate on small scales. On the contrary, if the star formation is a rapid cycle and GMCs are quickly destroyed by massive stars, then a decorrelation is expected at small scales between gas and young stars. In nearby galaxies, the latter is the phenomenon which has been clearly observed, i.e. the star-forming process is a rapid cycle at small scales (e.g. \citealt{schruba_scale_2010,kruijssen_fast_2019,chevance_lifecycle_020,kim_environmental_2022}).

Sub-kpc studies are challenging at higher redshifts (z) because of the fine resolution needed. One can take advantage of strong gravitational lensing to probe a target galaxy behind massive galaxies or galaxy clusters at increased spatial resolutions and magnified luminosities (e.g. \citealt{richard_locuss_2010,jones_resolved_2010,bayliss_probing_2014,livermore_resolved_2015,patricio_kinematics_2018}). Such background galaxies are often strongly stretched and sometimes show multiple images, so one needs to model the foreground mass distribution in order to reconstruct the shape of the target at a given redshift. This allows us to probe sub-kpc sizes, and in the most strongly lensed regions even scales < $\SI{100}{pc}$. Using this methodology, it is possible to resolve in galaxies at $z>1$ small-scale structures like star-forming clumps (e.g. \citealt{cava_nature_2018,messa_multiply_2022,claeyssens_star_2022}), giant molecular clouds (GMCs, e.g. \citealt{dessauges-zavadsky_molecular_2019,dessauges-zavadsky_molecular_2023}), or to make other measurements at sub-kpc scales, such as metallicity gradients (e.g. \citealt{patricio_resolved_2019}), kinematics (e.g. \citealt{girard_towards_2019}), or radial profiles (e.g. \citealt{nagy_radial_2022}).

In this paper, we investigate the rKS law in two strongly lensed galaxies at $z\sim 1$: the Cosmic Snake galaxy behind the galaxy cluster MACS J1206.2-0847, and A521-sys1, which we refer to as A521, behind the galaxy cluster Abell 0521. These two galaxies are typical main sequence (MS) star-forming galaxies at their redshifts, for which multi-wavelength observations are available from, in particular, the Hubble Space Telescope (HST) in several filters, and the Atacama Large Millimeter/submillimeter Array (ALMA).

The paper is structured as follows: in Sect. \ref{sec:observation} we present the HST and ALMA observations of the Cosmic Snake and A521 and their data reductions, as well as their gravitational lens modelling. In Sect. \ref{sec:methods} we present the measurements of $\Sigma \mathrm{SFR}$ and $\Sigma M_\mathrm{mol}$ in both galaxies. In Sect. \ref{sec:analysis} we analyse and discuss the integrated and resolved KS laws in the Cosmic Snake and A521. Finally, we give our conclusions in Sect. \ref{sec:conclusion}.

Throughout this paper, we adopt the $\Lambda$-CDM cosmology with $H_0 = \SI{70}{km.s^{-1}.Mpc^{-1}}$, $\Omega_{\mathrm{M}} = 0.3$, and $\Omega_{\Lambda} = 0.7$. We adopt the \citet{salpeter_luminosity_1955} initial mass function (IMF).

\section{Observations and data reduction}
\label{sec:observation}
    \subsection{Cosmic Snake and A521 galaxies}
    
\begin{figure}
        \centering
        \includegraphics[trim={2cm 3cm 2cm 3cm},width=0.48\textwidth]{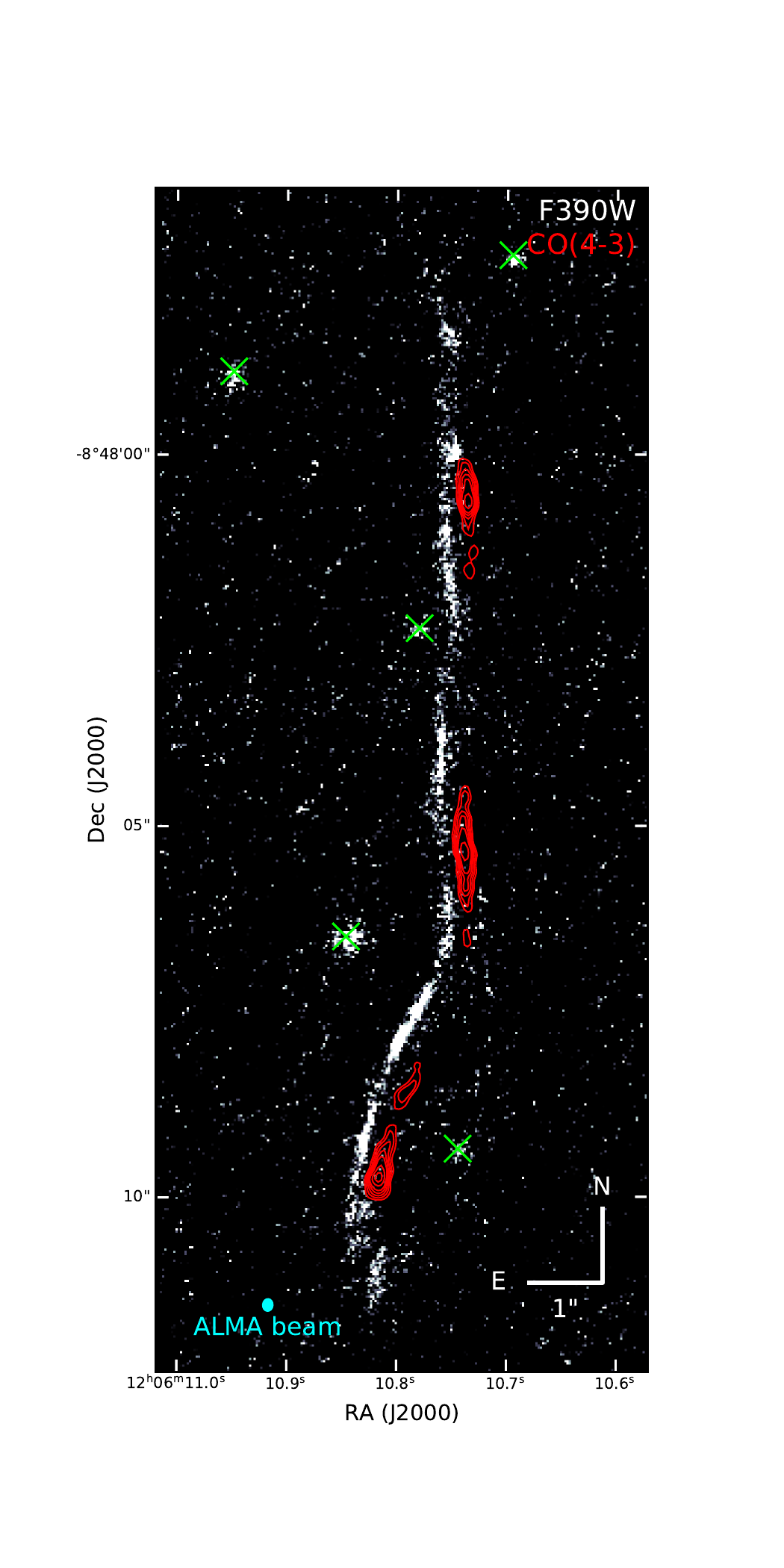}
        \caption{Rest-frame UV image with the F390W filter of HST of the Cosmic Snake galaxy's arc. The red contours correspond to the ALMA CO(4--3) velocity integrated intensity in levels of $4\mathrm{\sigma}$, $5\mathrm{\sigma}$, $6\mathrm{\sigma}$, $7\mathrm{\sigma}$, $8\mathrm{\sigma}$, $10\mathrm{\sigma}$, and $12\mathrm{\sigma}$ with an RMS noise of $\SI{0.020}{Jy.beam^{-1}.km.s^{-1}}$. The ALMA beam ($\SI{0.22}{\arcsec} \times \SI{0.18}{\arcsec}$) is displayed in blue. Green crosses indicate bright foreground sources.}
        \label{fig:CS_image}
    \end{figure}

\begin{figure}
        \centering
        \includegraphics[trim={0 0 0 0},width=0.48\textwidth]{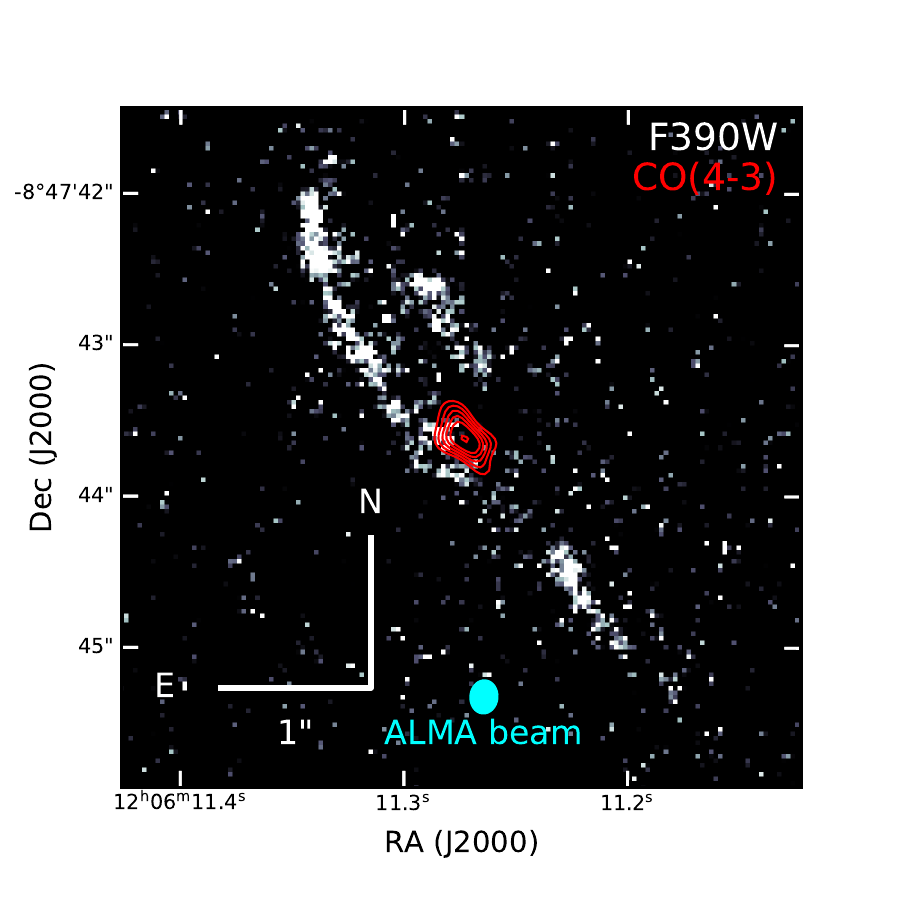}
        \caption{Rest-frame UV image with the F390W filter of HST of the Cosmic Snake galaxy's isolated counter-image. The red contours correspond to the ALMA CO(4--3) velocity integrated intensity in levels of $4\mathrm{\sigma}$, $5\mathrm{\sigma}$, $6\mathrm{\sigma}$, $7\mathrm{\sigma}$, $8\mathrm{\sigma}$, and $10\mathrm{\sigma}$ with an RMS noise of $\SI{0.025}{Jy.beam^{-1}.km.s^{-1}}$. The ALMA beam ($\SI{0.21}{\arcsec} \times \SI{0.18}{\arcsec}$) is displayed in blue.}
        \label{fig:CS_ci_image}
    \end{figure}

\begin{figure*}
        \centering
        \includegraphics[trim={1.8cm 2.6cm 1.8cm 2.8cm},width=0.98\textwidth]{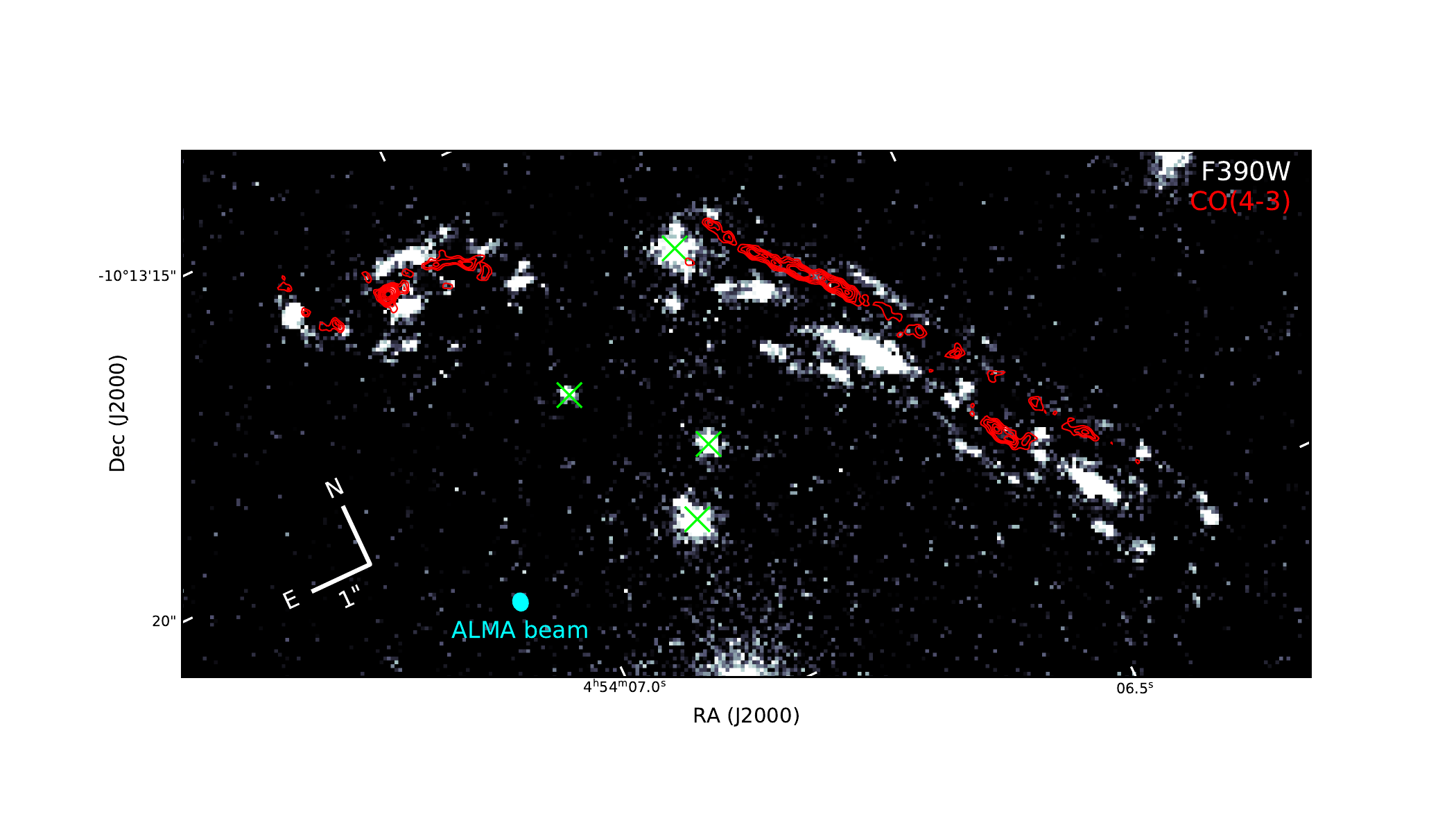}
        \caption{Rest-frame UV image with the F390W filter of HST of the A521. The red contours correspond to the ALMA CO(4--3) velocity integrated intensity in levels of $4\mathrm{\sigma}$, $5\mathrm{\sigma}$, $6\mathrm{\sigma}$, $7\mathrm{\sigma}$, $8\mathrm{\sigma}$, $10\mathrm{\sigma}$, $12\mathrm{\sigma}$, and $14\mathrm{\sigma}$ with an RMS noise of $\SI{0.010}{Jy.beam^{-1}.km.s^{-1}}$. The ALMA beam ($\SI{0.19}{\arcsec} \times \SI{0.16}{\arcsec}$) is displayed in blue. Green crosses indicate bright foreground sources.}
        \label{fig:A521_image}
    \end{figure*}

    The Cosmic Snake and A521 are two strongly lensed galaxies located behind the galaxy clusters MACS J1206.2-0847 and Abell 0521, respectively. They have several multiple images that are magnified by factors of a few to hundreds. For both of these galaxies we can see an arc including several images of the source galaxy with significant stretching and amplification, as well as an isolated counter-image with almost no stretching and amplification of a few (see Figs. \ref{fig:CS_image}, \ref{fig:CS_ci_image}, and \ref{fig:A521_image}). These galaxies are representative of MS star-forming galaxies at $z\sim 1$, with the Cosmic Snake having a stellar mass $M_{\star}=(4.0\pm 0.5)\times 10^{10}\,\si{M_\odot}$ and $\mathrm{SFR}=30\pm 10 \,\si{M_\odot.yr^{-1}}$, and A521 having $M_{\star}=(7.4\pm 1.2)\times 10^{10}\,\si{M_\odot}$ and $\mathrm{SFR}=26\pm 5 \,\si{M_\odot.yr^{-1}}$. More detailed descriptions of these galaxies can be found in \citet{patricio_kinematics_2018,patricio_resolved_2019}, \citet{girard_towards_2019}, \citet{nagy_radial_2022}, \citet{messa_multiply_2022}, and \citet{dessauges-zavadsky_molecular_2019,dessauges-zavadsky_molecular_2023}.

    \begin{table}
        \caption{Properties of the Cosmic Snake and A521}
        \label{table:properties}
        \centering                        
        \begin{tabular}{c c c}     
        \hline\hline                
            Quantity & Cosmic Snake & A521\\
        \hline                        
            $z$ & 1.036 & 1.044\\
            $M_{\star}$ & $(4.0\pm 0.5)\times 10^{10}\,\si{M_\odot}$ & $(7.4\pm 1.2)\times 10^{10}\,\si{M_\odot}$\\
            SFR & $30\pm 10 \,\si{M_\odot.yr^{-1}}$ & $26\pm 5\,\si{M_\odot.yr^{-1}}$\\
            $\Sigma \mathrm{SFR}$ & $1.5\pm \SI{0.1}{M_\odot.yr^{-1}.kpc^{-2}}$ & $1.8\pm \SI{0.1}{M_\odot.yr^{-1}.kpc^{-2}}$\\
            $\Sigma M_{\mathrm{mol}}$ & $570\pm \SI{60}{M_\odot.pc^{-2}}$ & $430\pm \SI{50}{M_\odot.pc^{-2}}$\\
        \hline
        \end{tabular}
        \tablefoot{The redshift values were derived from the CO(4--3) emission line observed with ALMA, for both galaxies. The values of $M_{\star}$ and SFR of the Cosmic Snake are taken from \citet{cava_nature_2018}, and those of A521 from \citet{nagy_radial_2022}. $\Sigma \mathrm{SFR}$ and $\Sigma M_{\mathrm{mol}}$ are derived in this paper for both galaxies, in Sect. \ref{sec:integratedKS}.}
    \end{table}

    \subsection{HST observations}

    We used the image of MACS J1206.2-0847 observed in F390W with WFC3/UVIS in the context of the Cluster Lensing And Supernova survey with Hubble (CLASH), as this filter corresponds to rest-frame ultraviolet (UV) wavelengths.\footnote{The data from CLASH are available at \url{https://archive.stsci.edu/prepds/clash/}.} The map we used has a point spread functions (PSF) resolution of $\SI{\sim 0.1}{^{\prime\prime}}$ and a pixel scale of $\SI{0.03}{^{\prime\prime}}$ \citep{cava_nature_2018}, and the exposure time was $\sim \SI{4959}{s}$. A full description of the CLASH dataset can be found in \citet{postman_cluster_2012}.
    
    We took the A521 images observed in F390W with WFC3/UVIS from the HST archive (ID: 15435, PI: Chisholm). The exposure time was $\SI{2470}{s}$. The software Multidrizzle \citep{koekemoer_cosmos_2007} was used to align and combine in a single image individual calibrated exposures. The final image has a PSF resolution of $\SI{0.097}{^{\prime\prime}}$ and a pixel scale of $\SI{0.06}{^{\prime\prime}}$ \citep{messa_multiply_2022}.

    \subsection{ALMA observations}
    
    The CO(4--3) emission of the Cosmic Snake was detected with ALMA in band 6 at $\SI{226.44}{GHz}$, corresponding to a redshift of $z=1.03620$. The observations were acquired in Cycle 3 (project 2013.1.01330.S), in the extended C38-5 configuration with a maximum baseline of $\SI{1.6}{km}$ and 38 antennas of the $\SI{12}{m}$ array. The total on-source integration time was $\SI{52.3}{min}$ \citep{dessauges-zavadsky_molecular_2019}. The isolated counter-image of the Cosmic Snake, as well as A521, were observed in band 6 in Cycle 4 (project 2016.1.00643.S), in the C40-6 configuration with a maximum baseline of $\SI{3.1}{km}$ and 41 antennas of the $\SI{12}{m}$ array. For the isolated counter-image of the Cosmic Snake, the total on-source time was $\SI{51.8}{min}$. For A521, it was $\SI{89.0}{min}$. The CO(4--3) line in A521 was detected at $\SI{225.66}{GHz}$, which corresponds to a redshift of $z=1.04356$ \citep{dessauges-zavadsky_molecular_2023}. The spectral resolution was set to $\SI{7.8125}{MHz}$ for all three observations.
    
    The data reduction was performed using the standard automated reduction procedure from the pipeline of the Common Astronomy Software Application (CASA) package \citep{mcmullin_casa_2007}. Briggs weighting was used to image the CO(4--3) emission with a robust factor of 0.5. Using the \emph{clean} routine in CASA interactively on all channels until convergence, the final synthesized beam size obtained for the Cosmic Snake galaxy was $\SI{0.22}{\arcsec} \times \SI{0.18}{\arcsec}$ with a position angle of $85^\circ$ for the arc, and $\SI{0.21}{\arcsec} \times \SI{0.18}{\arcsec}$ with an angle of $49^\circ$ for the isolated counter-image. For A521 the final synthesized beam size was $\SI{0.19}{\arcsec} \times \SI{0.16}{\arcsec}$ at $-74^\circ$. The adopted pixel scale for the CO(4--3) data cube is $\SI{0.04}{\arcsec}$ for the Cosmic Snake arc and $\SI{0.03}{\arcsec}$ for the Cosmic Snake isolated counter-image and A521. The achieved root mean squares (RMS) are $\SI{0.29}{mJy.beam^{-1}}$, $\SI{0.42}{mJy.beam^{-1}}$, and $\SI{0.20}{mJy.beam^{-1}}$, per $\SI{7.8125}{MHz}$ channel, for the Cosmic Snake arc, the Cosmic Snake isolated counter-image, and A521, respectively. The CO(4--3) moment-zero maps were obtained using the \emph{immoments} routine from CASA by integrating the flux over the total velocity range where CO(4--3) emission was detected.

\section{Methodology}
\label{sec:methods}

    \subsection{Gravitational lens model}

    The gravitational lens models used for the Cosmic Snake and A521 galaxies are constrained by multiple images found in HST observations. Lenstool \citep{jullo_bayesian_2007} was used to compute and optimise the models. The RMS accuracies of the lens models for the positions in the image plane of the Cosmic Snake and A521 galaxies are $\SI{0.15}{^{\prime\prime}}$ and $\SI{0.08}{^{\prime\prime}}$, respectively. More details on the gravitational lens models used for the Cosmic Snake and A521 can be found in \citet{cava_nature_2018} for the Cosmic Snake, and in \citet{richard_locuss_2010} and \citet{messa_multiply_2022} for A521.

    \subsection{Convolution}

    Since we compare quantities derived from ALMA and HST fluxes in small regions of the galaxies, we ensured that our HST and ALMA maps of a given galaxy are comparable by matching their resolutions. First we adjusted the pixel scale of the HST and ALMA images, then we convolved the HST images with the synthesised beam of the ALMA observations, and the ALMA images with the PSF of HST.


    \subsection{Determination of physical quantities}
    \label{subsec:physquant}
        \subsubsection{Molecular gas mass}
            We used the CO(4--3) line detected with ALMA as the tracer of $M_{\mathrm{mol}}$. First we converted the velocity-integrated flux of the CO(4--3) line ($S_{\mathrm{CO}}\Delta V$) into luminosity ($L'_{\mathrm{CO(4 \mbox{--} 3)}}$) using this equation from \citet{solomon_molecular_1997}:
            
            \begin{equation}
            \label{eq:LCO}
                L'_{\mathrm{CO(4 \mbox{--} 3)}} = 3.25 \times 10^{7} S_{\mathrm{CO(4 \mbox{--} 3)}}\Delta V\nu_{\mathrm{obs}}^{-2}D_{\mathrm{L}}^2(1+z)^{-3} (\si{K.km.s^{-1}.pc^2}),
            \end{equation}
            
            \noindent with $S_{\mathrm{CO}}\Delta V$ in $\si{Jy.km.s^{-1}}$, and where $\nu_{\mathrm{obs}}$ is the observed frequency in GHz, and $D_L$ the luminosity distance of the source in Mpc. The luminosity is then converted into $M_{\mathrm{mol}}$ \citep{dessauges-zavadsky_molecular_2019}:
            
            \begin{equation}
            \label{eq:Mmol}
                M_{\mathrm{mol}} = \left(\frac{\alpha_{\mathrm{CO}}}{\si{M_\odot(K. km. s^{-1}.pc^2)^{-1}}}\right)\left(\frac{L'_{\mathrm{CO(4 \mbox{--} 3)}}/0.33}{\si{K. km. s^{-1}.pc^2}}\right)\si{M_\odot},
            \end{equation}
            
            \noindent where we used the CO luminosity correction factor $r_{4,1}{=}L'_{\mathrm{CO(4 \mbox{--} 3)}}/L'_{\mathrm{CO(1 \mbox{--} 0)}}{=}0.33$, which was extrapolated from $r_{4,2}$ and $r_{2,1}$ measured in the Cosmic Snake \citep{dessauges-zavadsky_molecular_2019} and $z{\sim}1.5$ BzK galaxies \citep{daddi_co_2015}, respectively. We assumed the Milky Way CO-to-H$_2$ conversion factor $\alpha_{\mathrm{CO}}{=}\SI{4.36}{M_\odot(K. km. s^{-1}.pc^2)^{-1}}$, since both in the Cosmic Snake and A521 $\alpha_{\mathrm{CO}}$ was found to be close to the Milky Way value from the virialised mass of detected GMCs \citep{dessauges-zavadsky_molecular_2019,dessauges-zavadsky_molecular_2023}.

            In both galaxies, the CO(2--1) line was also detected with the Plateau de Bure Interferometer (PdBI) for the Cosmic Snake \citep{dessauges-zavadsky_molecular_2019}, and with the Institut de radioastronomie millimétrique (IRAM) $\SI{30}{m}$ single dish antenna for A521 \citep{dessauges-zavadsky_molecular_2023}. In both cases, the total molecular gas content traced by the CO(2--1) emission was identical to that traced by CO(4--3). We therefore conclude that using the CO(4--3) line to trace the molecular gas mass is reliable.

        \subsubsection{Star formation rate}
            
            We used the HST rest-frame UV observations with the F390W filter to compute SFR using Eq. (1) of \citet{kennicutt_star_1998}:
            
            \begin{equation}
            \label{eq:SFR}
                \mathrm{SFR} = 1.4\times 10^{-28} L_{\nu} (\si{ergs.s^{-1}.Hz^{-1}}),
            \end{equation}
            
            \noindent where $L_{\nu}$ is the UV luminosity. Furthermore, we applied an extinction correction as in \citet{calzetti_dust_2001} on the SFR as the UV continuum may be significantly affected by extinction:
            
            \begin{equation}
            \label{eq:exti}
                f_i(\lambda) = f_o(\lambda) 10^{0.4 E(B-V) k^e(\lambda)},
            \end{equation}
            
            \noindent with the obscuration curve for the stellar continuum $k^e(\lambda) = 1.17(-2.156+1.509/\lambda - 0.198/\lambda^2 + 0.011/\lambda^3)+1.78$ given by \citet{calzetti_dust_2000}, where $\lambda$ is the rest-frame wavelength in $\si{\mu m}$. The colour excess $E(B-V)$ was computed in \citet{nagy_radial_2022} both in radial bins and in the isolated counter-images of the galaxies by performing spectral energy distribution (SED) fitting on multiple HST bands. The values of $E(B-V)$ obtained from SED fits are in agreement with the value estimated from the Balmer decrement by \citet{messa_multiply_2022}. 


\section{Analysis and discussion}
\label{sec:analysis}
    \subsection{Integrated Kennicutt-Schmidt law}
        \label{sec:integratedKS}
        We measured the integrated $\Sigma M_{\mathrm{mol}}$ and $\Sigma \mathrm{SFR}$ on the isolated counter-image to the north-east of the arc for the Cosmic Snake, and on the counter-image to the east for A521. These counter-images show the entire galaxy for both galaxies, unlike the arcs where only a fraction of the galaxy is imaged. To compute $\Sigma M_{\mathrm{mol}}$ and $\Sigma \mathrm{SFR}$ we used the following method. We integrated both the CO(4--3) emission and the UV flux inside the half-light radius measured in the F160W band \citep{nagy_radial_2022}, then we converted them into the corresponding physical quantities $M_{\mathrm{mol}}$ (using Eqs. (\ref{eq:LCO}) and (\ref{eq:Mmol})) and SFR (using Eq. (\ref{eq:SFR})), by correcting the SFR for extinction using $E(B-V)$ computed in the same counter-images. To obtain $\Sigma M_{\mathrm{mol}}$ and $\Sigma \mathrm{SFR}$ we then divided by the respective half-light surfaces of the galaxies in the image plane. As both the fluxes and the surfaces were measured in the image plane, there is no need to correct for gravitational lensing if we assume a uniform amplification over the integration area. This is a fair assumption because the magnification varies only by $\sim 0.3$ and $\sim 0.5$ over the counter-images of the Cosmic Snake and A521, respectively.
        
        The uncertainty on $\Sigma M_{\mathrm{mol}}$ ($\Delta(\Sigma M_{\mathrm{mol}})$) was computed following
        
        \begin{equation}
        \label{eq:deltammol}
            \Delta(\Sigma M_{\mathrm{mol}}) = \frac{\Sigma M_{\mathrm{mol}}(\sigma_{\mathrm{RMS}})}{\sqrt{N_{\mathrm{pix,tot}}/N_{\mathrm{pix,beam}}}}
        \end{equation}
        
        \noindent where $\Sigma M_{\mathrm{mol}}(\sigma_{\mathrm{RMS}})$ is the root mean square noise ($\sigma_{\mathrm{RMS}}$) around the galaxy converted into $\Sigma M_{\mathrm{mol}}$ units, $N_{\mathrm{pix,tot}}$ is the total number of pixels inside the integration area, and $N_{\mathrm{pix,beam}}$ is the number of pixels in the beam. The uncertainty on $\Sigma \mathrm{SFR}$ ($\Delta(\Sigma \mathrm{SFR})$) was computed following
        
        \begin{equation}
        \label{eq:deltasfr}
            \Delta(\Sigma\mathrm{SFR}) = \sqrt{ \left( \frac{\Sigma\mathrm{SFR}(\sigma_{\mathrm{RMS}})}{\sqrt{N_{\mathrm{pix,tot}}/N_{\mathrm{pix,beam}}}}  \right)^2 + \left( \Sigma\mathrm{SFR}(\sigma_{\mathrm{phot}}) \right)^2 }
        \end{equation}

        \noindent where $\Sigma \mathrm{SFR}(\sigma_{\mathrm{RMS}})$ is the root mean square noise ($\sigma_{\mathrm{RMS}}$) around the galaxy converted into $\Sigma\mathrm{SFR}$, and $\Sigma\mathrm{SFR}(\sigma_{\mathrm{phot}})$ is the photometric error converted into $\Sigma\mathrm{SFR}$. We add the magnification uncertainties in quadrature, although these latter are negligible in comparison to other sources of uncertainties.
        
\begin{figure}
            \centering
            \includegraphics[width=0.48\textwidth]{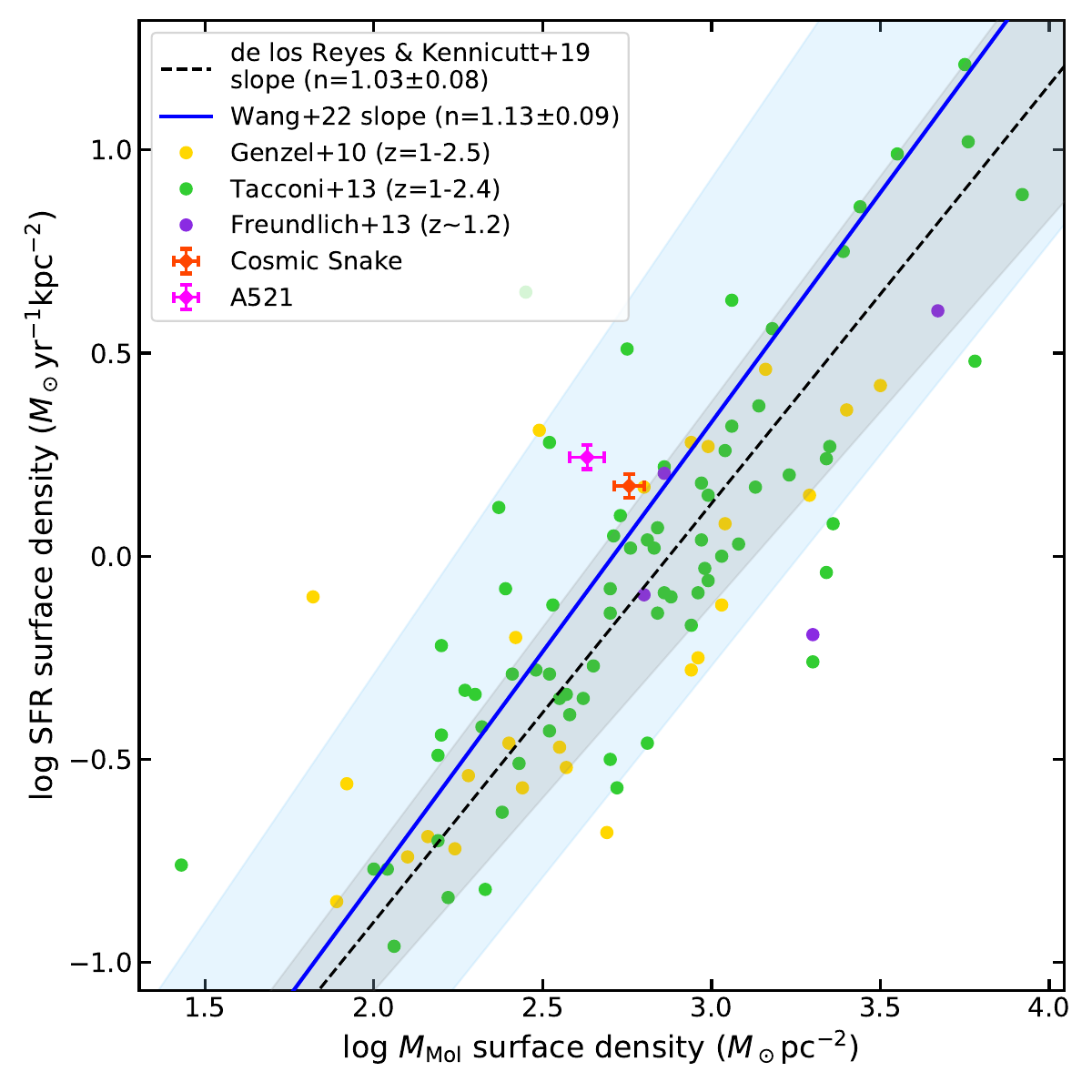}
            \caption{Integrated KS relation of the Cosmic Snake (red) and A521 (purple), along with a compilation of $z=1-2.5$ galaxies from \citet{genzel_study_2010} (yellow), \citet{tacconi_phibss_2013} (green), and \citet{freundlich_towards_2013} (violet). The blue line is the slope obtained by \citet{wang_3_2022} for MS SFGs at $z=0.4-3.6$. The dashed black line is the slope obtained for local galaxies by \citet{de_los_reyes_revisiting_2019}. The shaded areas indicate the uncertainties of the respective slopes.}
            \label{fig:integrated_KS}
        \end{figure}

        We find for the Cosmic Snake $\Sigma\mathrm{SFR} = 1.5\pm \SI{0.1}{M_\odot.yr^{-1}.kpc^{-2}}$ and $\Sigma M_{\mathrm{mol}} = 570\pm \SI{60}{M_\odot.pc^{-2}}$. For A521 we have $\Sigma\mathrm{SFR} = 1.8\pm \SI{0.1}{M_\odot.yr^{-1}.kpc^{-2}}$ and $\Sigma M_{\mathrm{mol}} = 430\pm \SI{50}{M_\odot.pc^{-2}}$. We show in Fig. \ref{fig:integrated_KS} the Cosmic Snake and A521 in the (integrated) KS diagram ($\Sigma \mathrm{SFR}-\Sigma M_{\mathrm{mol}}$), along with a compilations of 25 galaxies from \citet{genzel_study_2010} ($z=1-2.5$), 73 galaxies from \citet{tacconi_phibss_2013} ($z=1-2.4$), and 4 galaxies from \citet{freundlich_towards_2013} ($z\sim 1.2$). These galaxies are all MS star-forming galaxies (SFGs). We also plot the slope from \citet{de_los_reyes_revisiting_2019} for local spiral galaxies, as well as the slope obtained for stacks of MS SFGs by \citet{wang_3_2022} at $z=0.4-3.6$. The Cosmic Snake and A521 are clearly within the distribution of $z\gtrsim 1$ galaxies.

        Furthermore, the compilation of $z\gtrsim 1$ galaxies globally satisfies the KS relation with a slope of $1.13\pm 0.09$ \citep{wang_3_2022}, thus higher than $z=0$ galaxies. Such a steeper slope implies, for a given $\Sigma M_{\mathrm{mol}}$, a higher $\Sigma \mathrm{SFR}$ in distant galaxies than in the nearby ones. This might indicate that $z\gtrsim 1$ galaxies have higher star formation efficiencies. This is indeed the case, since the study of the integrated star formation efficiencies ($\mathrm{SFE} = M_{\mathrm{mol}}/\mathrm{SFR}$) of MS galaxies shows a mild increase of SFE with redshift \citep{tacconi_phibss_2018,dessauges-zavadsky_alpine-alma_2020,wang_3_2022}.
        

    \subsection{Resolved Kennicutt-Schmidt law}
    \label{sec:rKS}

        We studied the rKS law in different bin sizes in the Cosmic Snake and A521. To do so we created 6 grids paving the reconstructed source plane images of each galaxy, with boxes of $\SI{200}{pc}$, $\SI{400}{pc}$, $\SI{800}{pc}$, $\SI{1600}{pc}$, $\SI{2800}{pc}$, and $\SI{3200}{pc}$ for the Cosmic Snake, and $\SI{200}{pc}$, $\SI{400}{pc}$, $\SI{800}{pc}$, $\SI{1600}{pc}$, $\SI{3200}{pc}$, and $\SI{6400}{pc}$ for A521. We consider an additional larger bin size in A521 as the galaxy is more extended than the Cosmic Snake, with star formation happening up to a galactocentric radius of $\SI{8}{kpc}$ and molecular gas detected up to $\SI{6}{kpc}$, compared to, respectively, $\SI{7}{kpc}$ and $\SI{1.7}{kpc}$ in the Cosmic Snake \citep{nagy_radial_2022}. We then lensed the grids in the corresponding image plane. Due to the differential lensing, the area of some of these boxes is smaller than the matched PSF (HST PSF convolved with ALMA beam) in the image plane, so we discarded those boxes. This is the case for about half of the boxes of $\SI{200}{pc}$, and 20\% of the boxes of $\SI{400}{pc}$. We then measured $\Sigma M_{\mathrm{mol}}$ and $\Sigma \mathrm{SFR}$ inside each of the remaining boxes for both galaxies. In A521, one cluster member is present in front of a small part of the arc (corresponding to the upper green cross in Fig. \ref{fig:A521_image}). It has no significant diffuse emission so we simply masked it for our analysis.
        
        To estimate the flux inside a given box from the ALMA maps, we applied the technique developed for the Cosmic Snake galaxy in \citet{dessauges-zavadsky_molecular_2019} in the context of the search of molecular clouds. The method takes into account the three dimensions of the CO(4--3) datacube. To evaluate the detection threshold, the fidelity was computed as

        \begin{equation}
            \mathrm{fidelity}(\mathrm{S/N}) = 1- \frac{N_{\mathrm{neg}}(\mathrm{S/N})}{N_{\mathrm{pos}}(\mathrm{S/N})},
        \end{equation}

        \noindent where $N_{\mathrm{pos}}$ and $N_{\mathrm{neg}}$ are the number of positive and negative emission detections with a given signal-to-noise (S/N) in the primary beam, respectively \citep{walter_alma_2016,decarli_alma_2019}. The fidelity of 100\% was achieved at $\mathrm{S/N}=4.4$ in individual channel maps in both the Cosmic Snake and A521, and when considering co-spatial emission in two adjacent channels, it was reached at $\mathrm{S/N}=4.0$ in the Cosmic Snake and at $\mathrm{S/N}=3.6$ in A521. Therefore, in the ALMA datacube of the Cosmic Snake, we extracted for a given box the emission from each individual channel where the flux inside the box was above a $4.4\,\sigma$ RMS threshold, or above a $4.0\,\sigma$ threshold if the flux inside the same box in an adjacent channel was also above $4.0\,\sigma$. We did the same for the ALMA maps of A521 with thresholds of $4.4\,\sigma$ and $3.6\,\sigma$, respectively. Boxes below the ALMA RMS detection threshold are excluded, we do not consider upper limits. The HST flux is always detected where we detect CO. For each box we applied the extinction correction corresponding to the radial bin (computed in \citet{nagy_radial_2022}) where the majority of the pixels of the box lies. The uncertainties on $M_{\mathrm{mol}}$ and SFR were computed inside each box following Eqs. \ref{eq:deltammol} and \ref{eq:deltasfr}, respectively. 
        
\begin{figure*}
            \centering
            \includegraphics[trim={0 6cm 0 0},width=\textwidth]{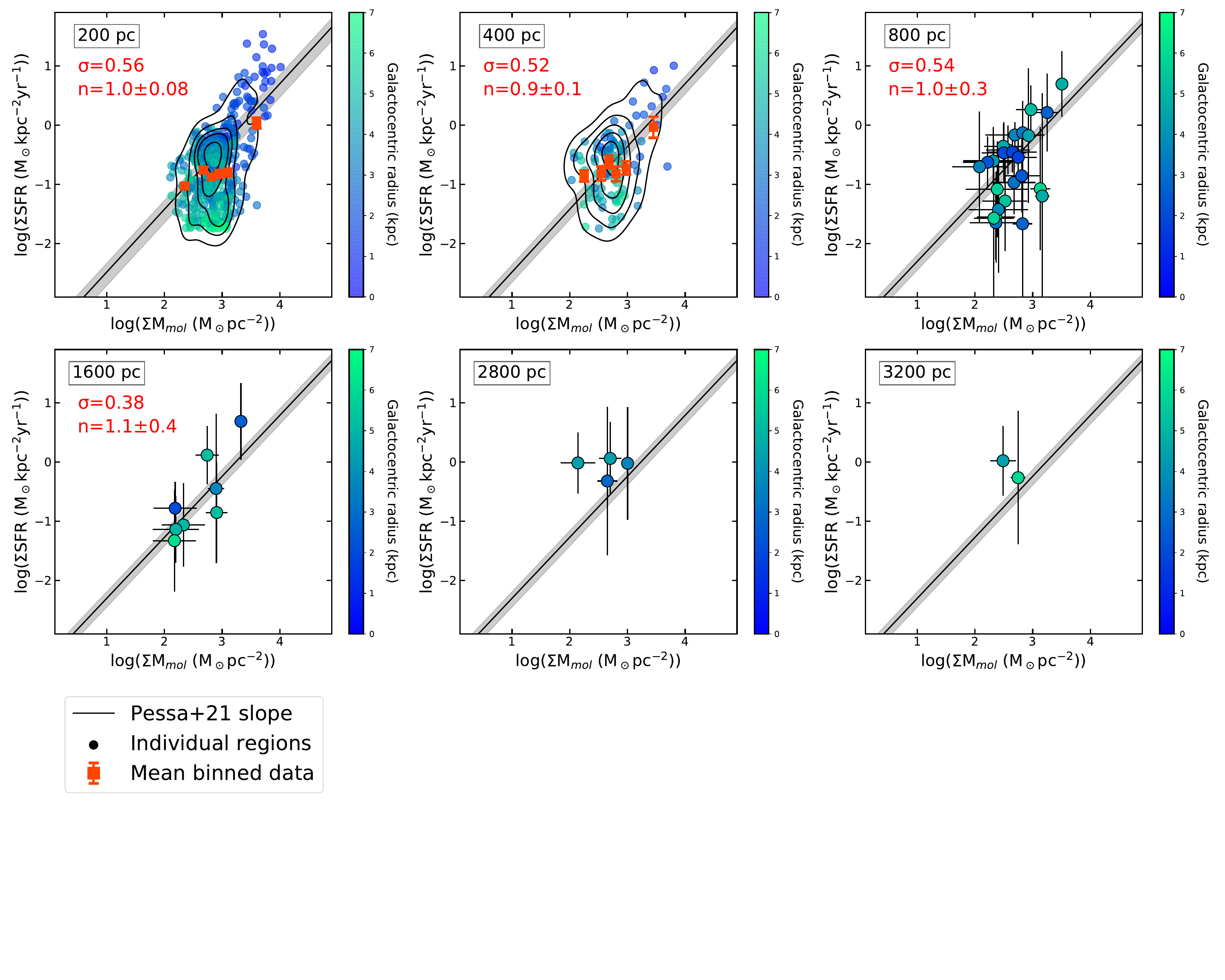}
            \caption{rKS diagram of the Cosmic Snake at different spatial scales. Each panel corresponds to a given bin size in the source plane, as indicated. The contours plotted in the first two panels are Kernel density estimates of the data. The five different contours correspond, from inside to outside to 10\%, 30\%, 50\%, 70\%, and 90\% iso-proportions of the density, respectively. The black line is the rKS slope obtained by \citet{pessa_star_2021} at the closest spatial scale to ours ($n_{\SI{100}{pc}} = 1.06$, $n_{\SI{500}{pc}} = 1.06$, and $n_{\SI{1}{kpc}} = 1.03$), with the grey shaded area showing the corresponding scatter ($\sigma_{\SI{100}{pc}} = 0.41$, $\sigma_{\SI{500}{pc}} = 0.33$, and $\sigma_{\SI{1}{kpc}} = 0.27$). The datapoints are coloured according to their galactocentric distance, as shown in the colour bars. The orange squares correspond to the $\Sigma \mathrm{SFR}$ means of datapoints within 6 $\Sigma M_{\mathrm{mol}}$ bins with equal number of datapoints. We report in the upper left corner the scatter ($\sigma$) of the datapoints with respect to the fits from \citet{pessa_star_2021}, as well as the slopes ($n$) obtained from a linear fitting for the bin sizes between $\SI{200}{pc}$ and $\SI{1600}{pc}$.}
            \label{fig:grid_snake}
        \end{figure*}

\begin{figure*}
            \centering
            \includegraphics[trim={0 8cm 0 0},width=\textwidth]{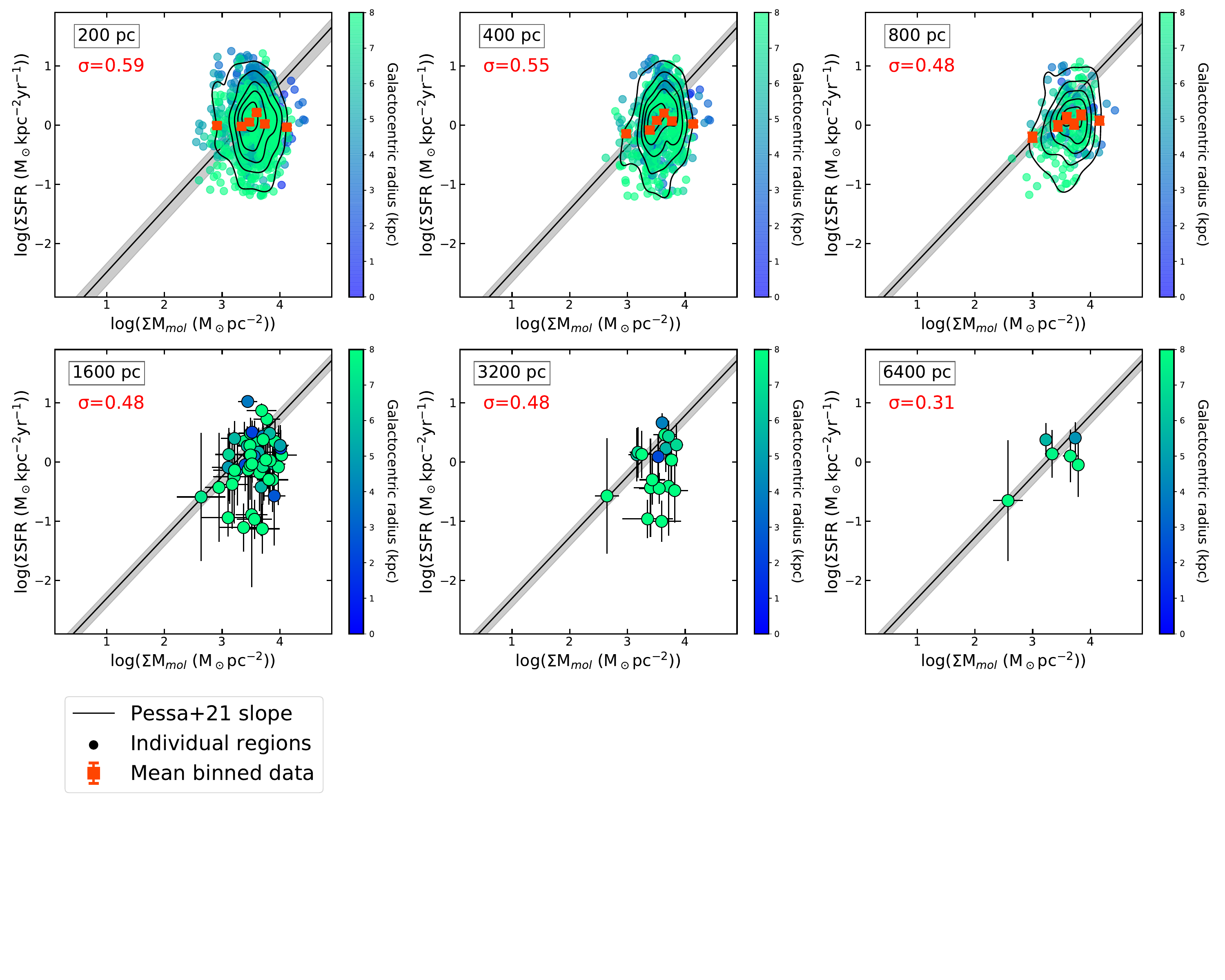}
            \caption{Same as Fig. \ref{fig:grid_snake} but for A521.}
            \label{fig:grid_A521}
        \end{figure*}

        The results for the rKS relation in the Cosmic Snake and A521 are displayed in Figs. \ref{fig:grid_snake} and \ref{fig:grid_A521}, respectively, with each panel corresponding to a different bin size. We display with orange squares the $\Sigma M_{\mathrm{mol}}$ values corresponding to the means in 6 x-axis bins  with equal number of datapoints. By performing a linear regression on all datapoints using the Levenberg-Marquardt algorithm\footnote{The algorithm takes into account the uncertainties on the $x$ and $y$ values of the datapoints} and least squares statistic in the Cosmic Snake for scales $\SI{\le 1600}{pc}$, we obtain slopes of $n^{\mathrm{CS}}_{\SI{200}{pc}} = 1.00\pm 0.08$, $n^{\mathrm{CS}}_{\SI{400}{pc}} = 0.9\pm 0.1$, $n^{\mathrm{CS}}_{\SI{800}{pc}} = 1.0\pm 0.3$, and $n^{\mathrm{CS}}_{\SI{1.6}{kpc}} = 1.1\pm 0.4$.  Uncertainties in the slope measurements increase with spatial scale due to the decrease of the number of datapoints. For the Cosmic Snake, the overall slope of the distribution for the bin sizes $\le\SI{1600}{pc}$ is similar to the slope reported for local galaxies. For larger scales ($\SI{> 1600}{pc}$) the number of boxes is too low for a reliable fit. For A521, no overall slope can be inferred at any bin size. The horizontal alignment of the binned means in Fig. \ref{fig:grid_A521} can be due to a lack of correlation in the data, as the binned means of a random distribution of points has the same horizontal alignment. We investigate below the differences between the two galaxies, and in particular, in the context of nearby samples from literature.
        
        In order to determine what is driving the difference in the distribution of datapoints between the two galaxies, we investigated the galactocentric effect, and used a colour coding depending on the galactocentric distance of each box. For the smaller scales of $\SI{200}{pc}$ and $\SI{400}{pc}$ we clearly see a segregation with the galactocentric distance in the Cosmic Snake galaxy. The boxes closer to the centre have much higher $\Sigma \mathrm{SFR}$ and $\Sigma M_{\mathrm{mol}}$ than the ones at large galactocentric radii. The Cosmic Snake has steep radial profiles of $\Sigma \mathrm{SFR}$ and $\Sigma M_{\mathrm{mol}}$ \citep{nagy_radial_2022}, so seeing a correlation between the galactocentric distance and the positions of the datapoints in the rKS diagram is not surprising. In A521, no segregation with the galactocentric distance is seen. This is in line with the shallow radial profiles of $\Sigma \mathrm{SFR}$ and $\Sigma M_{\mathrm{mol}}$ in A521, hence no significant difference in the rKS diagram between regions closer to the centre and regions in the outskirts is seen.

        \begin{figure*}
            \centering
            \includegraphics[trim={0 12cm 0 0},width=\textwidth]{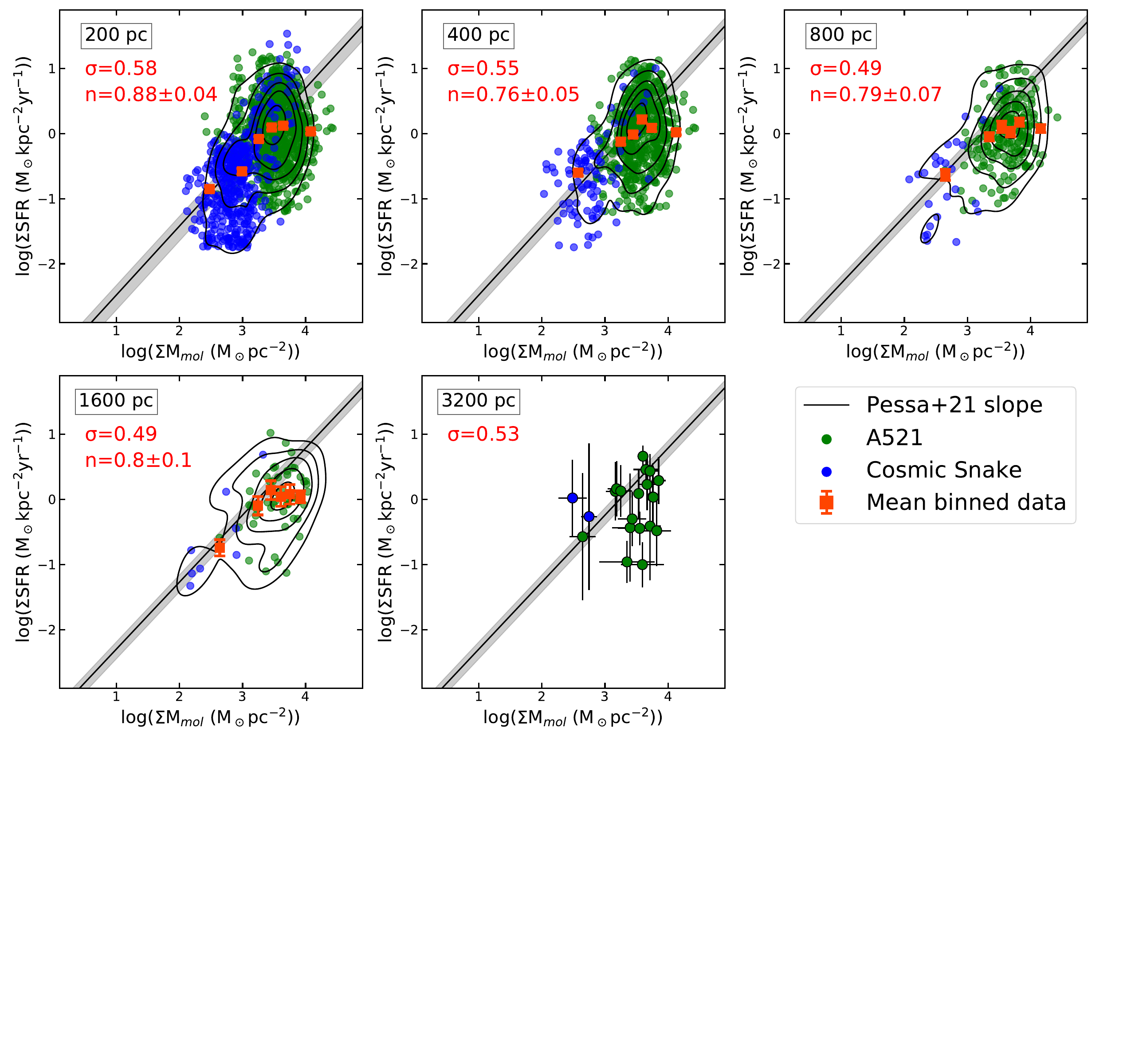}
            \caption{Same as Figs. \ref{fig:grid_snake} and \ref{fig:grid_A521} but for the combination of the Cosmic Snake and A521. The datapoints are no longer coloured according to galactocentric distance. Instead, the colour indicates the corresponding galaxy: green for A521 and blue for the Cosmic Snake.}
            \label{fig:grid_Both}
        \end{figure*}

        \citet{pessa_star_2021} measured the rKS in 18 star-forming galaxies from the Physics at High Angular resolution in Nearby GalaxieS (PHANGS\footnote{\url{https://phangs.org/}}) survey, at scales of $\SI{100}{pc}$, $\SI{500}{pc}$, and $\SI{1}{kpc}$. They reported slopes\footnote{The slopes from \citet{pessa_star_2021} have been computed by binning the x-axis and averaging the y-axis values inside each $\Sigma \mathrm{SFR}$ bin, whereas our slopes were computed by taking into account every measurement.} of $n_{\SI{100}{pc}} = 1.06\pm 0.01$, $n_{\SI{500}{pc}} = 1.06\pm 0.02$, and $n_{\SI{1}{kpc}} = 1.03\pm 0.02$, respectively, concluding that no evidence of systematic dependence on spatial scale is shown by the slopes. The slopes of local galaxies match those of the Cosmic Snake within error-bars, although our measurements have much bigger uncertainties due to sparser sampling.

        Moreover, in both galaxies, and specifically in A521, we lack dynamical range in $\Sigma \mathrm{SFR}$ and $\Sigma M_{\mathrm{mol}}$, especially for small values, to consistently constrain the rKS slope in $z\sim 1$ galaxies. $\Sigma \mathrm{SFR}$ spans $\sim 3.5$ orders of magnitude in the Cosmic Snake and $\sim 2.5$ in A521, compared to $\sim 5$ in the sample of 18 galaxies from \citet{pessa_star_2021}, and $\Sigma M_{\mathrm{mol}}$ spans $\sim 2$ orders of magnitude both in the Cosmic Snake and in A521, compared to $\sim 3$ in \citet{pessa_star_2021}. Higher sensitivity observations could allow to refine the estimation of the slope in the Cosmic Snake, or to enable to estimate the slope in A521. It is however important to note that the lack of dynamical range in A521 is not only due to a poor S/N, as the Cosmic Snake has a S/N comparable to A521 but a much better dynamical range.

        We plot the combined rKS of the Cosmic Snake and A521 in Fig. \ref{fig:grid_Both}, in order to increase the dynamical range of $\Sigma \mathrm{SFR}$ and $\Sigma M_{\mathrm{mol}}$. The slopes of the stacks ($n^{\mathrm{Stack}}$) are: $n^{\mathrm{Stack}}_{\SI{200}{pc}} = 0.88\pm 0.04$, $n^{\mathrm{Stack}}_{\SI{400}{pc}} = 0.76\pm 0.05$, $n^{\mathrm{Stack}}_{\SI{800}{pc}} = 0.79\pm 0.07$, and $n^{\mathrm{Stack}}_{\SI{1.6}{kpc}} = 0.8\pm 0.1$. These slopes are shallower than the slopes of the Cosmic Snake galaxy alone, and also than the slopes obtained by \citet{pessa_star_2021}. The reason is that A521 has a high density of points below the rKS line from \citet{pessa_star_2021}, as illustrated by the contours in Fig. \ref{fig:grid_A521}. One possible reason for these points to have so low SFR may be that the extinction is underestimated, and this, specifically, where the molecular gas density is high. It may also be due to the SFR tracer (rest-frame UV) we use, which traces star-forming regions with ages $\sim\SI{100}{Myr}$. For a long continuous star formation history (SFH), the estimated SFR by Eq. \ref{eq:SFR} would be accurate. However, in the case of a more bursty star formation with a constant SFH over a shorter time-frame of $\sim\SI{10}{Myr}$, Eq. \ref{eq:SFR} will underestimate the real SFR.


        For each set of datapoints at a given bin size, we compute the scatter in dex ($\sigma$) as the standard deviation of the datapoints around the rKS power law fits from \citet{pessa_star_2021} at the closest reported spatial scale (\SI{100}{pc}, \SI{500}{pc}, or \SI{1}{kpc}). We use this method instead of computing the scatter around the best-fitted power law like in \citet{pessa_star_2021} due to the uncertainty of the fit for the Cosmic Snake, and the meaningless fit if performed for A521. The values are reported in Table \ref{table:scatter}. Although the number of datapoints per grid binning size and the global shape of their distribution is notably different between the Cosmic Snake and A521, their respective scatters are similar at bin sizes up to $\SI{800}{pc}$. The scatter of both galaxies is also similar to the stack of the two, at those scales. At $\SI{1600}{pc}$, the scatter of the Cosmic Snake decreases significantly, whereas that of A521 stays constant up to $\SI{3200}{pc}$, then it decreases as well. The scatter decrease with increasing spatial scale is consistent with the results from \citet{bigiel_star_2008}, \citet{schruba_scale_2010}, and \citet{leroy_molecular_2013}. As a comparison, \citet{pessa_star_2021} reported scatters for the rKS law of $\sigma_{\SI{100}{pc}} = 0.41$, $\sigma_{\SI{500}{pc}} = 0.33$, and $\sigma_{\SI{1}{kpc}} = 0.27$. They argued that the decrease of scatter at increasing spatial scales is due to the averaging out of small scale variations.

        \begin{table*}
            \caption{Scatter ($\sigma$) of the rKS at spatial scales of $\SI{200}{pc}$, $\SI{400}{pc}$, $\SI{800}{pc}$, $\SI{1.6}{kpc}$, $\SI{3.2}{kpc}$, and $\SI{6.4}{kpc}$, computed as the standard deviation with respect to the fits from \citet{pessa_star_2021} at the closest spatial scale.}             
            \label{table:scatter}
            \centering                        
            \begin{tabular}{c c c c c c c}     
            \hline\hline                
                Galaxy & $\sigma_{\SI{200}{pc}}$ & $\sigma_{\SI{400}{pc}}$ & $\sigma_{\SI{800}{pc}}$ & $\sigma_{\SI{1600}{pc}}$ & $\sigma_{\SI{3200}{pc}}$ & $\sigma_{\SI{6400}{pc}}$\\
            \hline
                Cosmic Snake & 0.56 & 0.52 & 0.54 & 0.38 & & \\
                A521 & 0.59 & 0.55 & 0.48 & 0.48 & 0.48 & 0.31\\
                Stack of both & 0.58 & 0.55 & 0.49 & 0.49 & 0.53 & \\
            \hline
            \end{tabular}
        \end{table*}


    \subsection{Scale dependence of the $\Sigma \mathrm{SFR}$-$\Sigma M_\mathrm{mol}$ spatial correlation}
        
        We investigate the scale dependence of 
        the spatial correlation between $\Sigma \mathrm{SFR}$ and $\Sigma M_\mathrm{mol}$ in the Cosmic Snake and A521. As in \citet{schruba_scale_2010}, we do this by considering $\tau_{\mathrm{dep}} = \Sigma M_\mathrm{mol}/\Sigma \mathrm{SFR}$. $\tau_{\mathrm{dep}}$ is computed for apertures centred on CO and rest-frame UV peaks. The peaks were identified in the arcs of both galaxies, using the CO(4--3) emission from ALMA by \citet{dessauges-zavadsky_molecular_2019} for the Cosmic Snake and by \citet{dessauges-zavadsky_molecular_2023} for A521, and the rest-frame UV emission from HST by \citet{cava_nature_2018} for the Cosmic Snake and \citet{messa_multiply_2022} for A521. The CO peaks trace the GMCs, and the rest-frame UV peaks trace the star-forming regions. We then project the locations of the peaks in the source plane, and we centre apertures of different sizes on those positions. We use circular apertures with diameters of 200, 400, 800, 1200, and $\SI{1400}{pc}$ for the Cosmic Snake, and 400, 800, 1600, 3200, and $\SI{6400}{pc}$ for A521. These apertures are then lensed into the image plane, and we measure fluxes within each of them. As in Section \ref{sec:rKS}, we apply a $4.4\,\sigma$ RMS detection threshold to each individual channel of the ALMA datacubes for both the Cosmic Snake and A521, and a $4.0\,\sigma$ detection threshold in the case of co-spatial emission detections in two adjacent channels for the Cosmic Snake and $3.6\,\sigma$ for A521. Again, as the HST rest-frame UV emission is always detected where we also detect CO, we do not apply any detection threshold to the HST maps. We only consider apertures that are larger than the matched PSF in the image plane. We compute an average $\tau_{\mathrm{dep}}$ for each set of apertures of a given size and centred on a given type of peak (CO or UV). The uncertainty of a given average $\tau_{\mathrm{dep}}$ measurement ($\Delta\tau_{\mathrm{dep}}$) is computed as

        
        \begin{equation}
            \Delta\tau_{\mathrm{dep}} = \frac{\mathrm{std}(\tau_{\mathrm{dep}})}{\sqrt{N_{\mathrm{peaks}}}}
        \end{equation}
        
        \noindent where $\mathrm{std}(\tau_{\mathrm{dep}})$ is the standard deviation of all the $\tau_{\mathrm{dep}}$ used to compute the average, and $N_{\mathrm{peaks}}$ is the number of peaks.

\begin{figure*}
            \centering
            \begin{subfigure}{.49\textwidth}
                \includegraphics[width=\textwidth]{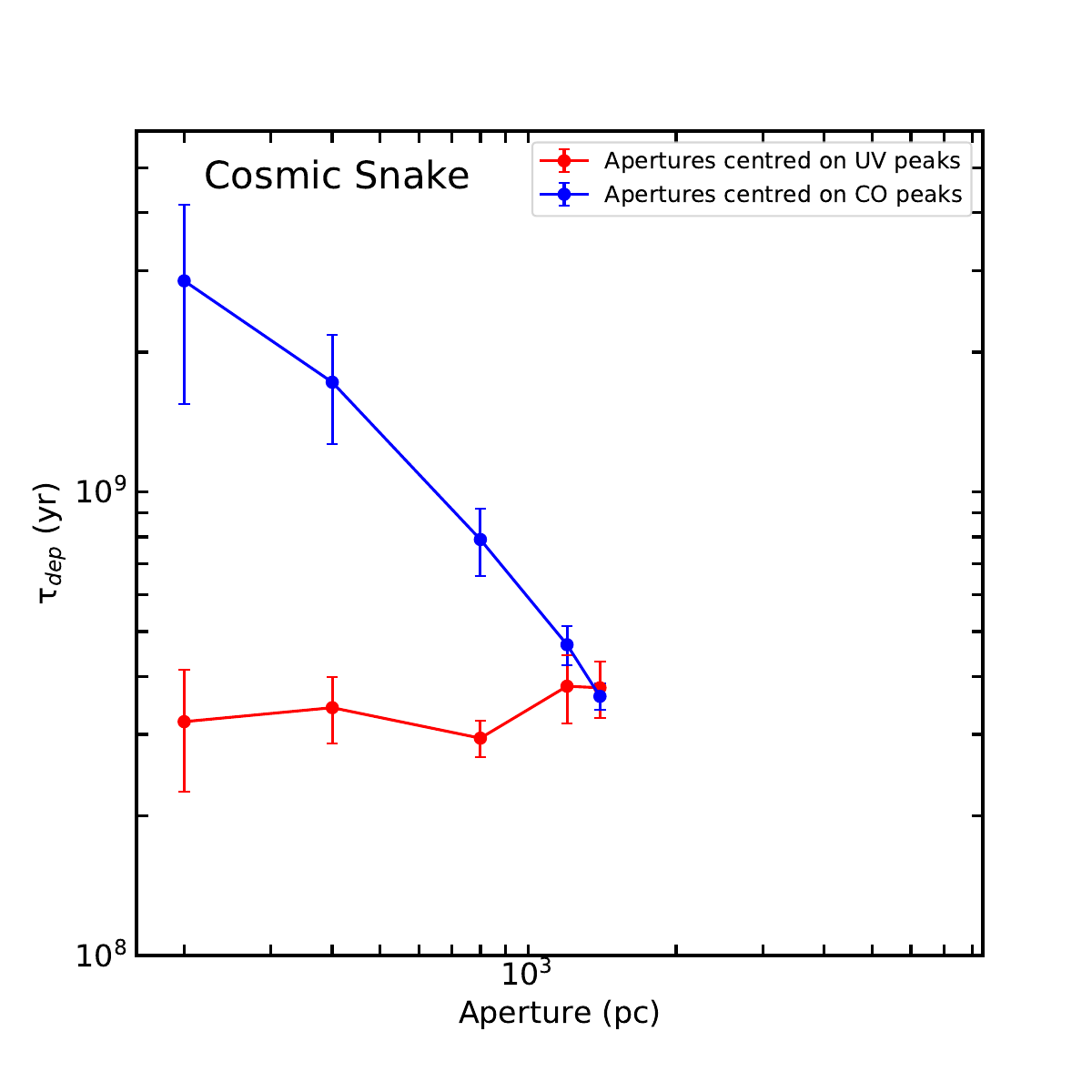}
            \end{subfigure}
            \begin{subfigure}{.49\textwidth}
                \includegraphics[width=\textwidth]{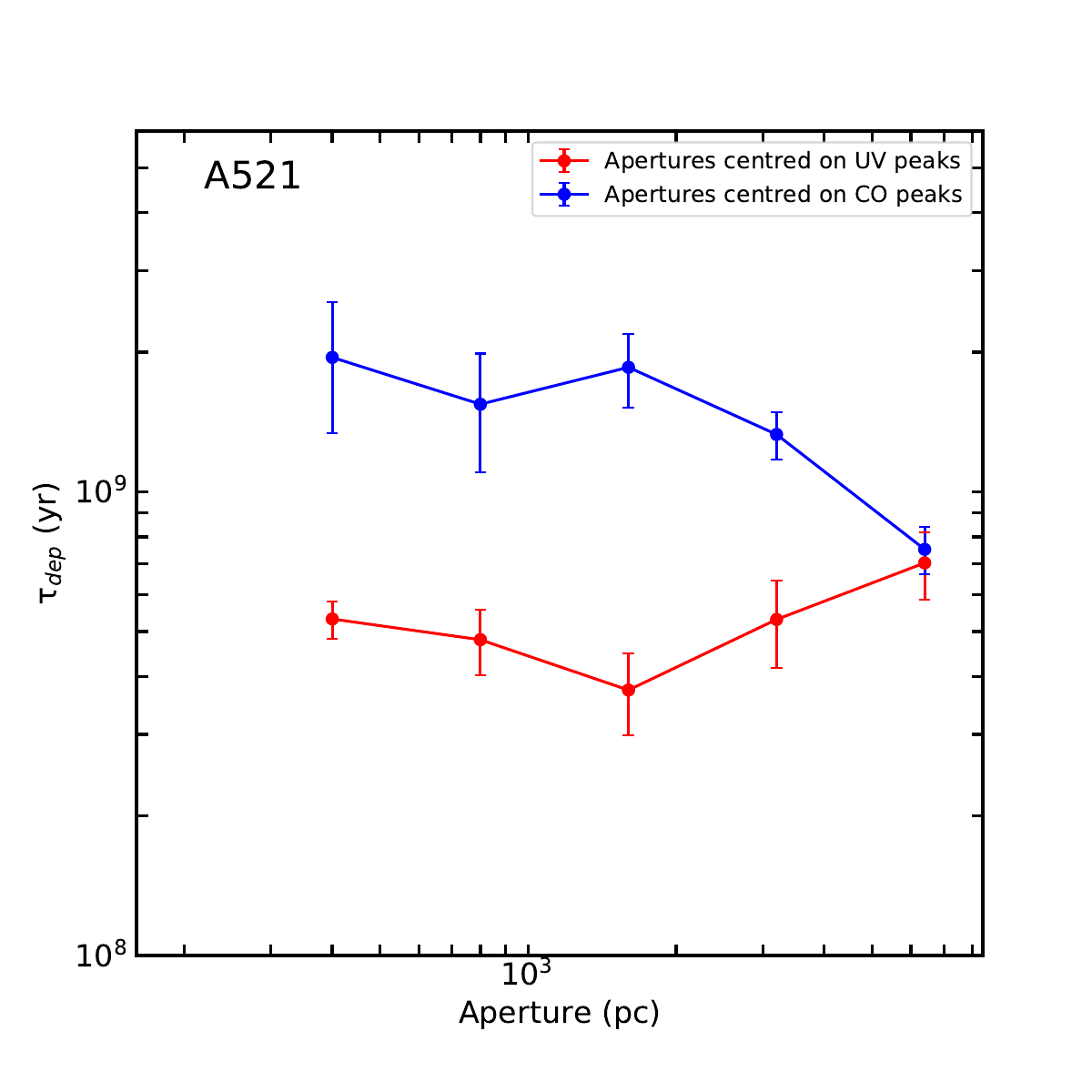}
            \end{subfigure}
            \caption{Scale dependence of the molecular gas depletion time in the Cosmic Snake (left) and A521 (right). The y-axis shows the average of $\tau_{\mathrm{dep}} = \Sigma M_\mathrm{mol}/\Sigma \mathrm{SFR}$ for different apertures, and centred on CO peaks (blue points) or on rest-frame UV peaks (red points). The diameters of the apertures are 200, 400, 800, 1200, and $\SI{1400}{pc}$ for the Cosmic Snake, and 400, 800, 1600, 3200, and $\SI{6400}{pc}$ for A521.}
            
            \label{fig:tuningfork}
        \end{figure*}

        The molecular gas depletion times for apertures of different sizes are given in Fig. \ref{fig:tuningfork}, showing separately the results for the apertures centred on CO peaks (blue points) and rest-frame UV peaks (red points). $\tau_{\mathrm{dep}}$ is strongly varying with the spatial scale (aperture size) and the type of emission targeted (CO or UV). From apertures larger than $\sim \SI{1}{kpc}$ in the Cosmic Snake and $\sim \SI{6}{kpc}$ in A521, the depletion times around CO peaks and the ones around rest-frame UV peaks are converging towards a common value.


        The overall behaviour of the molecular gas depletion time curves, commonly known as "tuning fork diagram", resembles that reported in the literature for local galaxies (e.g. \citealt{schruba_scale_2010,kruijssen_fast_2019,chevance_lifecycle_2020,kim_environmental_2022}). However, the uncertainties are much larger for the $z\sim 1$ galaxies because of the lack of statistics, as the number of detected clumps is about ten times lower than in a typical local galaxy. The $\tau_{\mathrm{dep}}$ convergence seems to happen at slightly larger scales in the Cosmic Snake ($\gtrsim \SI{1}{kpc}$), and much larger scales in A521 ($\sim \SI{6}{kpc}$), than in local galaxies ($\SI{500}{pc}-\SI{1}{kpc}$). Some plausible explanations for these differences are:
        \begin{itemize}
            \item The difference might be due to the difference of tracer, as local studies used $\mathrm{H\alpha}$ as the tracer of star-forming regions, but we used rest-frame UV emission which traces, on average, older star cluster complexes. As a result, the UV clumps that we detect are on average older ($\sim\SI{100}{Myr}$) than the $\mathrm{H\alpha}$ clumps ($\SI{10}{Myr}$) detected in nearby galaxies. This may imply that the dynamical drift is more significant because UV-bright star-forming regions have moved further away from their parent clouds for a given drift velocity.
            \item The drift of young stars from their parent molecular clouds might be faster in $z\sim 1$ galaxies than in local galaxies. This is expected from the larger gas fraction of high redshift galaxies, and also to their higher compactness; cloud-cloud collisions are enhanced and the gas is more dissipative, while the newly-formed stars are collisionless, and decouple faster from the gas. However, as argued in \citet{schruba_scale_2010} and \citet{chevance_lifecycle_2020}, the dynamical drift alone, at least in nearby galaxies, is not significant enough to be the cause of such large separations between GMCs and star-forming regions.
            \item Unless the stellar feedback is not strong enough, after $\SI{100}{Myr}$, the GMC parents of the UV clumps should already be destroyed if their lifetime is comparable to local GMCs ($10-\SI{30}{Myr}$; \citealt{kruijssen_fast_2019,chevance_lifecycle_2020}), explaining the lack of correspondence between the CO and UV peaks. Resolved $\mathrm{H\alpha}$ observations are needed to check how significantly the difference of tracers impacts the observed results.
            \item The star-forming regions detected in our high-redshift galaxies might not be born in the GMCs that we observe, but in other undetected clouds. In other words, there is no correspondence between the GMCs and the star-forming regions that we detect. In the Cosmic Snake and A521, small apertures only contain few peaks, and the majority of them is of the kind the aperture is centred on (CO or UV). Apertures of increasing sizes will include more and more peaks of both kinds, so the ratio of the CO peaks and UV peaks will converge towards 1. Therefore, the scale dependence of $\tau_{\mathrm{dep}}$ would actually trace the number of CO and UV peaks inside each aperture. 
        \end{itemize}

        An explanation of the increasing scatter at smaller spatial scales seen in the rKS plots (Figs. \ref{fig:grid_snake}, \ref{fig:grid_A521}, and \ref{fig:grid_Both}) and discussed in Sect. \ref{sec:rKS} may be found in the divergence at small scales of the molecular gas depletion time curves. At large scales ($>\SI{1}{kpc}$ in the Cosmic Snake and $>\SI{6}{kpc}$ in A521), any aperture chosen results in proportional fluxes of molecular gas and SFR tracers, even when focusing specifically at either star-forming regions or GMCs. This means that at these large scales, a randomly selected aperture will likely have a $\Sigma\mathrm{SFR}$ and a $\Sigma M_{\mathrm{mol}}$ which satisfy the rKS relation, consequently, the scatter of the relation for a sample of randomly selected apertures larger than $\SI{1}{kpc}$ (Cosmic Snake) or $\SI{6}{kpc}$ (A521) will be low, which is what we observe (Table \ref{table:scatter}). However, as apertures get smaller, there is much larger scatter because individual star-forming regions are at different stages of time evolution and have thus different CO-to-UV ratios. Focusing on for example a GMC results in a large $\tau_{\mathrm{dep}}$ because the flux from the tracer of SFR is missed, and $\tau_{\mathrm{dep}}$ is dominated by the numerator $\Sigma M_{\mathrm{mol}}$ (and vice-versa). This is the reason of the divergence at small spatial scales seen in Fig. \ref{fig:tuningfork}. However, when using a random gridding with a small bin size like in the $\Sigma\mathrm{SFR}$-$\Sigma M_{\mathrm{mol}}$ plots, the boxes happen to be sometimes between a CO peak and a rest-UV peak, resulting in a datapoint which satisfies the rKS relation, but sometimes they also fall right on a given peak, which yields a datapoint with either a high $\Sigma\mathrm{SFR}$ and a low $\Sigma M_{\mathrm{mol}}$, or the opposite. This is the cause of the large scatter seen at small scales in the rKS diagrams. As a result, the majority of the datapoints do not satisfy the rKS law at small scales, but the entire cloud of datapoints is centred on it, and even the slope is close to the value obtained for scales $>\SI{1}{kpc}$ (in the case of the Cosmic Snake). If the rKS law was valid at small scales, any randomly selected aperture would fall on the slope of the relation, within the scatter observed for the largest scales.

\section{Conclusions}
\label{sec:conclusion}
    We analysed the KS law in the Cosmic Snake and A521, two strongly lensed galaxies at $z\sim 1$, at galactic integrated scales down to sub-kpc scales. We used the rest-frame UV emission from HST to trace SFR and the CO(4--3) emission line detected with ALMA to trace $M_{\mathrm{mol}}$. In addition to several multiple images with magnifications of $\mu>20$ which are significantly stretched, and where only a fraction of the galaxy is visible, both galaxies show an isolated counter-image with overall uniform magnifications of 4.3 and 3 for the Cosmic Snake and A521, respectively. In those counter-images, the entirety of the galaxies is visible, thus we used them to compute integrated values of SFR and $M_{\mathrm{mol}}$. We found $\Sigma\mathrm{SFR} = 1.5\pm \SI{0.1}{M_\odot.yr^{-1}.kpc^{-2}}$ and $\Sigma M_{\mathrm{mol}} = 570\pm \SI{60}{M_\odot.pc^{-2}}$ in the Cosmic Snake, and $\Sigma\mathrm{SFR} = 1.8\pm \SI{0.1}{M_\odot.yr^{-1}.kpc^{-2}}$ and $\Sigma M_{\mathrm{mol}} = 430\pm \SI{50}{M_\odot.pc^{-2}}$ in A521. The two galaxies satisfy the integrated KS relation derived at $z=1-2.5$ \citep{genzel_study_2010,tacconi_phibss_2013,freundlich_towards_2013,wang_3_2022}.

    To study the rKS law by taking advantage of the strong gravitational lensing in the Cosmic Snake and A521, we defined 6 different grids in the source plane of each galaxies. We then lensed those grids in the image plane, and computed $\Sigma M_{\mathrm{mol}}$ and $\Sigma \mathrm{SFR}$ inside each box. The grids that we used had sizes of $\SI{200}{pc}$, $\SI{400}{pc}$, $\SI{800}{pc}$, $\SI{1600}{pc}$, $\SI{2800}{pc}$, and $\SI{3200}{pc}$ for the Cosmic Snake, and $\SI{200}{pc}$, $\SI{400}{pc}$, $\SI{800}{pc}$, $\SI{1600}{pc}$, $\SI{3200}{pc}$, and $\SI{6400}{pc}$ for A521.

    We derived the following results from the analysis of the rKS law in the Cosmic Snake and A521:
    \begin{itemize}
            \item We were able to perform a linear regression on the measurements in the Cosmic Snake for scales $\SI{\le 1600}{pc}$, obtaining slopes of $n^{\mathrm{CS}}_{\SI{200}{pc}} = 1.00\pm 0.08$, $n^{\mathrm{CS}}_{\SI{400}{pc}} = 0.9\pm 0.1$, $n^{\mathrm{CS}}_{\SI{800}{pc}} = 1.0\pm 0.3$, and $n^{\mathrm{CS}}_{\SI{1.6}{kpc}} = 1.1\pm 0.4$. These slopes are similar to those typically found in local galaxies. For A521 no overall slope could be inferred at any scale. We measured slopes for the combined rKS of the Cosmic Snake and A521 of $n^{\mathrm{Stack}}_{\SI{200}{pc}} = 0.88\pm 0.04$, $n^{\mathrm{Stack}}_{\SI{400}{pc}} = 0.76\pm 0.05$, $n^{\mathrm{Stack}}_{\SI{800}{pc}} = 0.79\pm 0.07$, and $n^{\mathrm{Stack}}_{\SI{1.6}{kpc}} = 0.8\pm 0.1$.
            \item To consistently constrain the rKS slopes in the analysed $z\sim 1$ galaxies, we lack dynamical range in both $\Sigma M_{\mathrm{mol}}$ and $\Sigma \mathrm{SFR}$. In the study of 18 PHANGS galaxies from \citet{pessa_star_2021}, both quantities span at least 1 more order of magnitude than our study.
            \item We see a clear spatial segregation in the distribution of the datapoints in the rKS diagram of the Cosmic Snake. Points close to the galactic centre tend to have higher $\Sigma M_{\mathrm{mol}}$ and $\Sigma \mathrm{SFR}$, whereas measurements in the outskirts show lower values. No such segregation is observed in A521. These observations match with the results from \citet{nagy_radial_2022} showing that the Cosmic Snake has much steeper radial profiles than A521, in $\Sigma M_{\mathrm{mol}}$ and $\Sigma \mathrm{SFR}$ in particular.
            \item The scatter of the datapoints in the Cosmic Snake and A521 is very similar at small scales up to $\SI{800}{pc}$. The scatter of both galaxies decreases at higher scales of $\SI{1600}{pc}$ for the Cosmic Snake, and $\SI{6400}{pc}$ for A521. The decrease of scatter at increasing bin sizes is similar to what is observed in $z=0$ galaxies and is due to the averaging out of small scale variations.
    \end{itemize}
    
    We measured the average $\tau_{\mathrm{dep}}$ inside apertures of different diameters centred on either rest-frame UV or CO(4--3) emission peaks in the Cosmic Snake and A521. In both galaxies, we observe the same overall behaviour as in local galaxies, that is the $\tau_{\mathrm{dep}}$ values measured using small apertures are clearly different whether the apertures are centred on rest-frame UV peaks or on CO(4--3) peaks, and they converge towards a common value at spatial scales large enough. In nearby galaxies, the convergence typically happens in apertures of diameters of $\SI{500}{pc}-\SI{1}{kpc}$. In the Cosmic Snake the $\tau_{\mathrm{dep}}$ measurements converge at higher but comparable apertures of size $\gtrsim \SI{1}{kpc}$, whereas in A521 it happens in much larger apertures of $\sim \SI{6}{kpc}$.

    We conclude that the increasing scatter in the rKS diagrams in small bin sizes is partly explained by the divergence observed between $\tau_{\mathrm{dep}}$ measured when focusing on rest-frame UV peaks and CO(4--3) peaks at small scales. By taking values of $\Sigma\mathrm{SFR}$ and $\Sigma M_{\mathrm{mol}}$ from randomly selected boxes of small sizes, the corresponding datapoint may satisfy the rKS law, but may also be significantly off if the aperture happens to capture only one kind of peak. In boxes larger than the size for which the $\tau_{\mathrm{dep}}$ values converge, any datapoint will tend to fall on the rKS, hence the smaller scatter. In the Cosmic Snake and A521, the scales at which $\tau_{\mathrm{dep}}$ converges are the scales at which the scatter of those galaxies in the rKS diagram decreases.

\begin{acknowledgements}
    This work was supported by the Swiss National Science Foundation.
    
    Based on observations made with the NASA/ESA Hubble Space Telescope, and obtained from the Data Archive at the Space Telescope Science Institute, which is operated by the Association of Universities for Research in Astronomy, Inc., under NASA contract NAS 5-26555. These observations are associated with program \#15435.
    
    This paper makes use of the following ALMA data: ADS/JAO.ALMA\#2013.1.01330.S, and ADS/JAO.ALMA\#2016.1.00643.S. ALMA is a partnership of ESO (representing its member states), NSF (USA) and NINS (Japan), together with NRC (Canada), MOST and ASIAA (Taiwan), and KASI (Republic of Korea), in cooperation with the Republic of Chile. The Joint ALMA Observatory is operated by ESO, AUI/NRAO and NAOJ.
    
    M.M. acknowledges the support of the Swedish Research Council, Vetenskapsrådet (internationell postdok grant 2019-00502).

    JS acknowledges support by the Natural Sciences and Engineering Research Council of Canada (NSERC) through a Canadian Institute for Theoretical Astrophysics (CITA) National Fellowship.
\end{acknowledgements}

%
%

\bibliographystyle{aa} 
\bibliography{references.bib}

\begin{appendix}

\end{appendix}

\end{document}